\documentclass{ws-procs9x6}
\usepackage{graphicx}

\begin{document} 

\title{
Quantum Monte Carlo calculations of symmetric nuclear matter
} 

\author{S. GANDOLFI and F. PEDERIVA}
\address{Dipartimento di Fisica and INFN, University of Trento,
via Sommarive 14, I-38050 Povo, Trento Italy}
\author{S. FANTONI}
\address{S.I.S.S.A., International School of Advanced Studies and INFN
via Beirut 2/4, 34014 Trieste, Italy}
\author{K. E. Schmidt}
\address{Department of Physics, Arizona State University, Tempe, AZ, USA}

\begin{abstract}
We present an accurate numerical study of the equation of state
of nuclear matter
based on realistic nucleon--nucleon interactions by means of Auxiliary Field Diffusion Monte Carlo (AFDMC)
calculations. The AFDMC method samples the spin and isospin degrees of freedom
allowing for quantum simulations of large nucleonic systems 
and represents an important step forward towards a quantitative understanding of 
problems in nuclear structure and astrophysics. 
\end{abstract}

\bodymatter

\section{INTRODUCTION}

The knowledge of the properties of 
symmetric and  asymmetric nuclear matter, is one of the main ingredients
of the equations predicting the structure, the dynamics and
the evolution of stars, in particular during their last stages, when they become
ultra--dense neutron stars. The large variety of available results,
partly due to the use of many different approximations, and 
partly to the limited knowledge of the nucleon--nucleon interaction,
lead to different interpretations and hypothesis. Depending on the
model used, the density of nuclear matter in the inner shells can reach 
up to 9 times the core density of stable nuclei, 
$\rho_0$ = 0.16 fm$^{-3}$\cite{piekarewicz04}, or be limited to 
4-5 times $\rho_0$.

One important step towards the understanding of these astrophysical 
problems is the study of the symmetric nuclear matter, related to the 
various model NN interactions available.
While considerable advances have been made\cite{morales02,akmal98},
it is still impossible to understand how different results depend
on the model interaction used rather than on the particular
approximation employed for computing energies.


Therefore, it is not possible to draw conclusions when comparing with
astronomical observations.
However, the recent success in predicting the 
properties of light nuclei gives us some confidence
that the non--relativistic description of nuclear matter based on effective 
potentials fitted to reproduce NN data and the binding energy of
light nuclei can be reliable enough.
The main feature of such nucleon--nucleon interactions, besides the short range repulsion,
is the explicit dependence on the relative quantum state of the nucleons which can be
described using spin and isospin, angular momentum, spin--orbit, and
tensor operators\cite{pandharipande79}. 

Properties of light nuclei (A$\le$6) can be efficiently computed with high 
accuracy using modern few--body techniques\cite{kamada01,gazit06,carlson98} or 
with the ab initio no-core nuclear shell model (with A$\le$12\cite{navratil00}).
Quantum Monte Carlo techniques
based on recasting the Schroedinger equation into a diffusion
equation (Diffusion Monte Carlo or Green's Function Monte Carlo),
allowed for performing calculations up to A$\le$12\cite{pieper05,pieper02}. 
However, the computational resources needed for such simulations
are very large, because of the summation
over all the possible states necessary to evaluate the terms of the
Hamiltonian with a quadratic dependence on spin and isospin. The number of 
such terms, and the CPU time needed to calculate them,  grows exponentially
with the number of nucleons; $^{12}$C\cite{pieper05} or 14 neutrons\cite{carlson03} is the limit for 
the currently available computational resources.

Since the spatial degrees of freedom are already sampled,
one would like to replace 
the sum over the spin--isospin states with an efficient sampling method.
The simulation of symmetric nuclear matter requires a minimum of 28 nucleons
in a box replicated in space (7 nucleons for each spin--isospin state) to
obtain a wave function with closed shells of momenta, and this is out of the reach of
the standard Quantum Monte Carlo methods.

Recently\cite{gandolfi06b} we shown that the spin--isospin is efficiently sampled
by using the Auxiliary Field Diffusion Monte Carlo method\cite{schmidt99}, 
which is based on the use of auxiliary variables to linearize the quadratic
spin--isospin operators of the nuclear matter Hamiltonian, making them treatable 
in a diffusion Monte Carlo scheme. Up to now it has been applied to simulate
pure neutron matter (up to 114 neutrons)\cite{fantoni01,sarsa03}, and 
neutron drops\cite{gandolfi06,pederiva04} interacting
with realistic two-- plus three--body interactions.

Here we extend calculations
to include isospin degrees of freedom and to deal with
the strong tensor--isospin force, responsible of the nuclear binding.
The method can readily handle an
asymmetry in the number of neutron and protons or the deformation of 
heavy nuclei.


\section{METHOD}

Auxiliary Field Diffusion Monte Carlo (AFDMC)\cite{schmidt99} is an extension 
of the standard Diffusion Monte Carlo method in which the ground state
of an Hamiltonian $H$ is obtained by solving the imaginary time dependent
Schroedinger equation. 
The solution is obtained by evolving a population of configurations 
of the system ("walkers") $X=\{R,\sigma,\tau\}$, where 
$R=\{\vec{r}_1,\dots,\vec{r}_N,\}$, 
$\sigma=\{\vec{\sigma}_1,\dots,\vec{\sigma}_N\}$, and $\tau=\{\vec{\tau}_1,\dots,\vec{\tau}_N\}$,
with $F(X,t)=\Psi_T(X)\Psi(X,t)$, according to
\begin{equation}
F(X,t)=\int dX'
\frac{\Psi_T(X)}{\Psi_T(X')} G_0(X,X',t) F(X',0)
\end{equation} 

The function  $\Psi_T$ is a ``trial'' wave function, usually determined by means
of variational calculations, and $G_0$ is  an approximation to 
the Green's function of the imaginary time Schroedinger equation:
\begin{equation}
G_0(X,X',t)=(4\pi D t)^{-3A/2}e^{-(R-R')^2/4Dt}e^{-t(V(X)-E_0)},
\end{equation} 
where $D=\hbar^2/2m$, $E_0$ is an estimate of the ground state energy of the 
system, and $V(X)$ is the nucleon--nucleon interaction.
For a long enough imaginary time the distribution of the walkers converges
to the product $\Psi_T(X)\Psi_0(X)$ where $\Psi_0$ is the wave function of
the ground state of $H$. This fact allows the  computation of matrix elements
$\langle \Psi_T|\hat O |\Psi_0\rangle$ of 
observables $\hat O$ of interest in a Monte Carlo way. When $\hat O\equiv \hat H$
the value obtained is the exact ground state energy of the system.

The presence of an interaction $V(X)$ including
operators like $(3\vec{\sigma}_i\cdot\hat{r}_{ij}\vec{\sigma}_j\cdot\hat{r}_{ij}-
\vec{\sigma}_i\cdot\vec{\sigma}_j)$ and $\vec{\tau}_i\cdot\vec{\tau}_j$ is the origin
of the computational cost in the standard approaches. 
AFDMC uses the
Hubbard--Stratonovich method to transform the operators $\hat S$ which are quadratic in 
the spin and isospin into linear operators:
\begin{equation}
e^{-(1/2)t\lambda\hat S^2} = \frac{1}{\sqrt{2\pi}}\int_{-\infty}^{+\infty}
dye^{-y^2/2}e^{y\sqrt{-\lambda t}\hat S}
\end{equation}

Then $\hat S$ are operators which are linear combinations of the
spin and isospin operators for each nucleon, and $\lambda$
depend on the interaction. The transformed Green's Function is applied to 
the spin-isospin part of the wave function, and its effect consists 
of a rotation of the spin and isospin degrees of freedom (written as four-component spinors
in the proton-neutron up-down basis)
by a quantity that depends on the auxiliary variable $y$ along with multiplication
of the state by an overall factor. The 
sum over spin and isospin is replaced by sampling a set of rotations of 
the variables.
This procedure reduces the dependence of the computational time on the number of
nucleons necessary
for performing a simulation step from exponential
to cubic. It is therefore possible to perform on a regular
workstation or on a modest PC cluster calculations
that would require Tflop supercomputers with the standard methods. 
This method, like other diffusion Monte Carlo methods,
suffers from the so--called ``sign problem''
when it is applied to fermions, and when complex wave functions need to be 
used. In our calculations we apply the fixed-phase approximation to 
overcome this problem\cite{zhang03}. While this method has already been successfully 
applied to pure neutron matter\cite{sarsa03}, it has not been previously
used for mixed proton and neutron systems. It should be noted that it
does not guarantee an upperbound to the mixed energy used here.
As a test for the correctness and the efficiency of our approach we
reproduced within 0.3 MeV total energy the
existing results for the binding energy of $^4$He with potentials of 
the AV6 class\cite{wiringa02}. We have also been able to compute
binding energies for $^{16}$O and for $^{40}$Ca\cite{inprep} with this method. 

\section{INTERACTION}
A crucial point in dealing with nuclear matter is 
the choice of the interaction among nucleons. As already mentioned, several
modern two--body potentials are available nowadays, all fitting the NN  
data with $\chi^2\sim 1$. We use
the potentials of the Argonne class with $n$ operators
(AVn)\cite{wiringa95}. While the full version contains $n=18$ operators,
most of the physics is reasonably well described by the first
6 operators made up of 4  central spin--isospin dependent components 
and two tensor ones, which include the long range
one--pion--exchange force. 
The most important missing terms are the
spin--orbit components. In nuclei and neutron drops the spin--orbit contribution to the
energy amounts to a few
tenths of MeV/nucleon. A correction of the  order 1MeV/nucleon can be attributed
at low densities to the remaining terms included in AV18. 
Specifically, we have used the interaction Argonne $v_8'$\cite{wiringa02}
truncated by dropping the spin--orbit terms, and including only the  
first six operators, which we denote as  AV6'.

It is well known that two--body NN interaction underbind light nuclei, and one needs to add a specific
effective three--body potential to reproduce their low energy properties.
Semi--phenomenological three--nucleon interactions following the lowest order three--nucleon diagrams
with one and two intermediate Delta resonance states provide a very satisfactory description of the
ground state energy and the low level spectra of light nuclei up to $^{12}$C\cite{pieper05}.
We have disregarded such three--body forces in our simulations.
In nuclear matter they are essential to reproduce the experimental saturation density, and, in general, 
they contribute about $10\%$ of the total binding energy. A full comparison with the available
experimental data goes beyond the scopes of the present paper. Here we are interested in showing 
the efficiency of the AFDMC methods in dealing with nuclear matter models which include 
realistic tensor interactions like in the AV6' potential. 


\section{RESULTS}
The results of the calculations with A=28 include box corrections that have computed by 
adding to the two body sums contribution of nucleons in the first shell of periodic cells.
Such procedure is effective. In order to assess the magnitude of finite size effects we 
performed calculations with 76 and 108 nucleons at densities $\rho$ = 0.08 fm$^{-3}$ 
and $\rho$ = 0.48 fm$^{-3}$. The results\cite{gandolfi06b} coincide with the ones obtained with 28 nucleons 
within 3 percent. 

We computed the EOS 
of symmetric nuclear matter in the range of densities 
$0.5\leq(\rho/\rho_0)\leq 3$, and compared it with previous available
results obtained with the same potential using
Fermi Hypernetted Chain in the Single Operator Chain approximation
(FHNC/SOC)  and the 
Brueckner-Hartree-Fock (BHF) in the two--hole line approximation\cite{bombaci05}. 
AFDMC calculations were performed with 28 
nucleons, filling the shell of plane waves with momentum of
modulus 1 and providing a wave function yielding an isotropic density.

\begin{figure}
\psfig{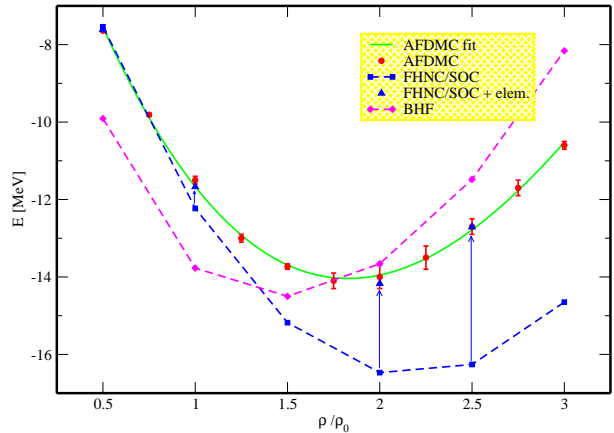}
\\
\caption{(color online).
Equation of state of symmetric nuclear matter calculated with different methods. Red circles 
represent AFDMC results with statistical error bars and the green line is the fitted functional form 
described in the text. 
Dashed lines correspond to calculations performed with other methods\cite{bombaci05}
(blue line with squares: FHNC/SOC; magenta with diamonds: BHF). Blue triangles represent 
the FHNC/SOC energies corrected by including the low order of elementary diagrams as described 
in the text. Blue arrows show the corresponding energy shift, which increases at higher densities.}
\end{figure}

The results are summarized in Fig. 1 and reported in Ref. \refcite{gandolfi06b}. The comparison of the various EOS
suggests the following comments: 
FHNC/SOC leads to an overbinding at high density.
A similar indication was found by Moroni et al.\cite{moroni95} after a DMC calculation 
of the EOS of normal liquid $^3$He
at zero temperature, with a guiding function including triplet and backflow correlations.
The comparison with the equivalent FHNC/SOC calculations of Refs. \refcite{manousakis83,viviani88} 
have shown similar discrepancies.
There are two main intrinsic approximations in variational FHNC/SOC calculations, which violate
the variational principle. The first one consists in neglecting a whole class of 
cluster diagrams, the so called {\sl elementary diagrams}, which cannot be summed up 
by means of FHNC integral equations. We have calculated the lowest order diagram of this class,
namely the one having only one correlation bond and four exchange bonds.
The results obtained show a substantial effect from this diagram and
bring the FHNC/SOC estimates very close to AFDMC results, as shown in Fig. 1.
The second approximation is related to the non-commutativity of the correlation operators
entering the variational wave function. The only class of cluster diagrams contributing
to such non-commuting terms, which can be realistically calculated, is that characterized by single
operator chains. It is believed that such an approximation is reliable in
nuclear matter, but there is no clear proof of this.
 
 
BHF calculations of ref.\cite{bombaci05} predict an EOS with a shallower binding
than the AFDMC one. It has been shown for symmetric nuclear matter, 
using the AV18 and AV14 potentials,
that contributions from three hole--line diagrams add a repulsive contribution up to $\sim$ 3MeV 
at densities below $\rho_0$\cite{song98}, and decrease the energy at high 
densities\cite{baldo01}. Such corrections, if computed with the AV6'
potential, would probably preserve the same general behavior, and bring the BHF EOS 
closer to the AFDMC one. Therefore, our 
calculations show that the two hole--line approximation used in Ref. \refcite{bombaci05}
is too poor, particularly at high density.

The AFDMC equation of state was fitted with the following functional form:
\begin{equation}
\frac{E}{A} = \frac{E_0}{A} + \alpha (x-\bar{x})^2 + \beta (x-\bar{x})^3 ,
\end{equation}
where $x=\rho/\rho_0$ and the various
coefficients are given by $E_0/A$ = -14.04(4) MeV, $\alpha$ = 3.09(6) MeV, 
$\beta$ = -0.44(8) MeV, and  $\bar{x}$ = 1.83(1). The resulting compressibility ${\it K} = 
9\bar{x}^2\left(\partial^2\left(E/A\right)/\partial x^2\right)_{\bar{x}}$ 
at saturation density $\bar{x}$ is $\sim$ 190 MeV.
The fit of the EOS  allows for computing the pressure vs. density for 
symmetric nuclear matter. 

\section{CONCLUSIONS}
The availability of an efficient and relatively fast projection algorithm for the computation of energies and 
other observables of dense hadronic matter enables the possibility of a more quantitative
understanding of the properties of neutron stars and supernovae, as well as that of medium--heavy nuclei. Computations
on such systems are at present out of reach of the standard GFMC methods and available 
supercomputers. Therefore, the extension of AFDMC algorithm to deal with nuclear matter is
a significant step forward. Some technical improvements on the calculations 
presented here, such as the addition to the
AV6' of spin--orbit terms and three--body interactions are 
already underway. The treatment of asymmetric nuclear matter, particularly important
for the determination of the properties of neutron stars, is also straightforward,
and will be the subject of future exploration.

We acknowledge helpful conversations with M.H. Kalos, 
P. Faccioli, W. Leidemann, E. Lipparini, and G. Orlandini.
This work was in part supported by NSF grant PHY-0456609. Calculations were performed on the 
HPC facility "BEN" at ECT* in Trento under a grant for Supercomputing Projects.

\bibliographystyle{ws-procs9x6}
\bibliography{cortona06}

\end{document}